\documentstyle[twocolumn,aps,epsf]{revtex}
\def\k{{\bf{k}}}

\def\q{{\bf{q}}}

\def\vereq#1#2{\lower3pt\vbox{\baselineskip1.5pt \lineskip1.5pt
\ialign{$#1\hfill##\hfil$\crcr#2\crcr\sim\crcr}}}

\begin{document}

\title{Phase Diagram of the Hubbard Model: Beyond the Dynamical Mean Field}
\author{M.~Jarrell$^1$, Th.~Maier$^2$, M.~H.~Hettler$^3$, A.N.~Tahvildarzadeh$^1$}
\address{$^1$ Department of Physics, University of Cincinnati, Cincinnati,
OH 45221}
\address{$^2$ Institut f\"ur Theoretische Physik, Universit\"at Regensburg,
93040 Regensburg\\} 
\address{$^3$ Forschungszentrum Karlsruhe, Postfach 3640, 76021 Karlsruhe, 
Germany}
\date{\today}
\maketitle

\begin{abstract}
        The Dynamical Cluster Approximation (DCA) is used to 
study non-local corrections to the dynamical mean field phase 
diagram of the two-dimensional Hubbard model. Regions of 
antiferromagnetic, d-wave superconducting, pseudo-gapped non-Fermi 
liquid, and Fermi liquid behaviors are found, in rough agreement 
with the generic phase diagram of the cuprates.  The non-local 
fluctuations beyond the mean field both suppress the 
antiferromagnetism and mediate the superconductivity.     
\end{abstract}

\paragraph*{Introduction}
The rich phenomenology of high-$T_c$ superconductors\cite{htc_review} 
has stimulated strong experimental and theoretical interest in the field of 
strongly correlated electron systems. Common to all high-$T_c$ systems is 
the presence of antiferromagnetic ordering in undoped samples in proximity 
to a superconducting phase with a $d$-wave order parameter and the normal 
state pseudogap dominating the physics in underdoped samples.  A successful
theory must describe all these fundamental features at the same time.

The 2D Hubbard model in the intermediate coupling regime or closely 
related models like the t-J model are believed to capture the essential 
physics of the high-$T_c$ cuprates \cite{Anderson}. The antiferromagnetic 
phase of the cuprates is well understood. In the strong coupling limit 
$U \gg W$, where $U$ is the Coulomb repulsion and $W$ the bare bandwidth, 
the undoped Hubbard model reduces to the Heisenberg model, which has been 
proven to describe the low energy spin fluctuations of the cuprate parent
compounds. However, off half-filling there is no complete understanding of 
the superconducting phase or the normal state pseudogap in the intermediate 
coupling 2D Hubbard model.

Finite size quantum Monte Carlo (QMC) calculations for the doped 2D Hubbard 
model in the intermediate coupling regime with $U\alt W$, support the idea 
of a spin fluctuation driven interaction mediating $d$-wave 
superconductivity \cite{scalapino1}.  However, the fermion sign problem and 
the fact that the number of degrees of freedom grows rapidly with the lattice 
size, limits these calculations to temperatures too high to study a possible 
transition\cite{scalapino1}.  These calculations are also restricted to 
relatively small system sizes making statements for the thermodynamic 
limit problematic, and inhibiting studies of the low energy physics.

These shortcomings do not apply to the Dynamical Mean Field Approximation 
(DMFA), which is by construction in the thermodynamic limit. Unfortunately, 
the lack of non-local dynamics in the DMFA inhibits a possible transition 
to a state with a non-local ($d$-wave) order parameter. 

However, it is well known within phenomenological theories that 
short-ranged antiferromagnetic spin fluctuations mediate pairing with 
$d$-wave symmetry and cause a pseudogap in underdoped samples
\cite{scalapino1,Monthoux,timusk}. There is experimental evidence that 
the correlation length of dynamical spin fluctuations in the optimally 
doped cuprates is very small\cite{zha}. 
Therefore, microscopic studies, which account for short-ranged dynamical 
correlations in addition to the local correlations of the DMFA are relevant 
and might succeed in describing the physics of the cuprates.     

The recently developed Dynamical Cluster Approximation 
(DCA)\cite{DCA_hettler,DCA_maier,DCA_huscroft} is a fully causal approach 
which systematically incorporates these non-local corrections to the DMFA 
by mapping the lattice problem onto an embedded periodic cluster of size 
$N_c$. For $N_c=1$ the DCA is equivalent to the DMFA and by increasing 
$N_c$ the dynamic correlation length can be gradually increased while the 
calculation remains in the thermodynamic limit. Previous DCA calculations 
have indicated the presence of an extended d-wave superconducting phase 
at moderate doping.\cite{DCA_maier,DCA_huscroft,DCA_lich}

In this manuscript we present calculations of the full phase diagram of 
the 2D Hubbard model studied with the DCA. To solve the cluster problem we use 
QMC. We choose $N_c=4$, the smallest cluster that includes non-local 
corrections while preserving the full translational and point group 
symmetries of the lattice. The results are compared to DMFA calculations, 
$N_c=1$, to illustrate the effect of the initial non-local corrections.  

\paragraph*{Formalism}  
A detailed discussion of the DCA formalism was given in previous 
publications \cite{DCA_hettler,DCA_maier,DCA_huscroft,DCA_moukouri0}.
The compact part of the free energy is coarse-grained in reciprocal 
space, projecting it onto a finite-sized cluster of $N_c$ points embedded 
in a self-consistently determined host.  The cluster problem is solved 
using the Hirsch-Fye impurity algorithm\cite{hirsch_fye} modified to 
simulate an embedded cluster\cite{DCA_moukouri0}, and the spectra are
analytically continued with the maximum entropy method\cite{JARRELLandGUB}.
Once the cluster problem has been solved, lattice susceptibilities may be 
calculated\cite{DCA_hettler}.  All calculations are done in the normal, 
paramagnetic state. We search for continuous phase transitions 
indicated by the divergence of the corresponding susceptibilities.  

\begin{figure}[htb]
\epsfxsize=3.3in
\epsffile{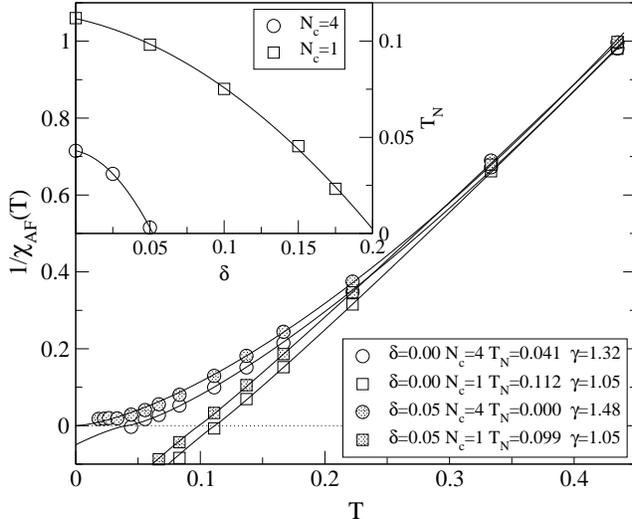}
\caption{Inverse antiferromagnetic susceptibility versus temperature 
for $U=2$.  The lines are fits to the function 
$1/\chi_{AF}(T) = b(T-T_N)^\gamma$.  For $N_c=1$
$\gamma\approx 1$, the mean-field value.  For $N_c=4$ non-local 
fluctuations suppress the transition, so that $\gamma$ increases 
and $T_N$ decreases (see inset).}
\label{Chi_AF}
\end{figure}
\begin{figure}[htb]
\epsfxsize=3.3in
\epsffile{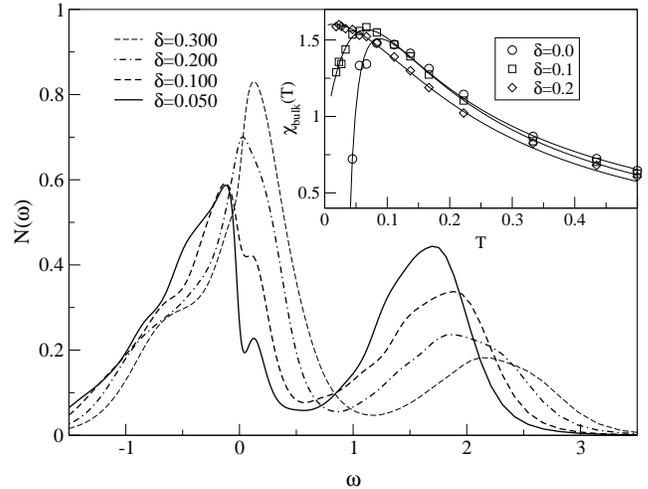}
\caption[a]{{The single-particle density of states $N(\omega)$
for $U=2$, $T=0.023$ and $N_c=4$. 
The inset shows the bulk susceptibility as a function of temperature.  
For $\delta < 0.2$ a peak develops at $T=T^*$ accompanied  by the evolution 
of a  pseudogap in the DOS for $T<T^*$.}}
\label{DOS_CHI}
\end{figure}
\begin{figure}[htb]
\epsfxsize=3.3in
\epsffile{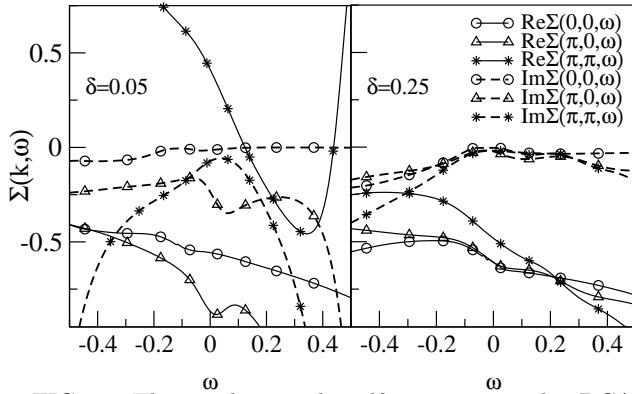}
\caption{The single-particle self energies at the DCA cluster momenta, 
plotted versus frequency for $U=2$, $T=0.023$, $N_c=4$ and $\delta=0.05$ (left)
and $\delta=0.25$ (right).
The self energy at ($\pi$,0) changes from non-Fermi liquid like at doping 
$\delta=0.05$ to Fermi liquid like at $\delta=0.25$.
}
\label{Sigmas_comp}
\end{figure}
\begin{figure}[htb]
\epsfxsize=3.3in
\epsffile{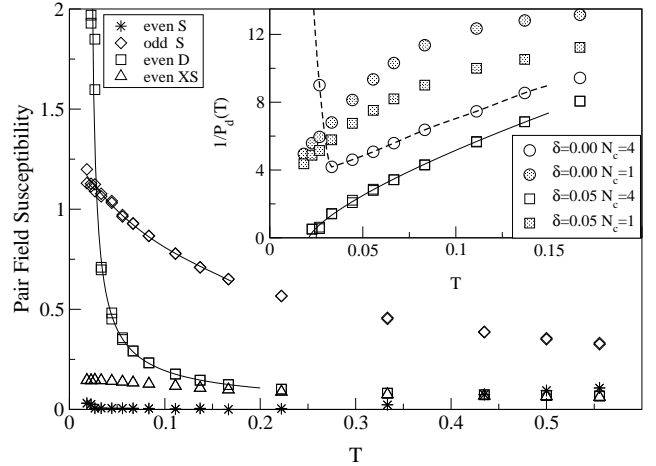}
\caption{The s-wave, extended s-wave, and d-wave even frequency and 
the odd-frequency s-wave $\q=0$ susceptibilities versus temperature 
for $U=2.0$, $\delta=0.05$, and $N_c=4$.  Pairing is found only in the 
even-frequency $\q=0$ d-wave channel. In the inset the 
inverse d-wave pair-field susceptibility is plotted versus temperature 
for two different dopings and cluster sizes.  The line is a fit to 
$1/P_{d}(T) = b(T-T_c)^\gamma$ with $T_c=0.021$ and $\gamma=0.72$.}
\label{Pairing}
\end{figure}

The Hubbard model is characterized by a near-neighbor hopping $t$ and an
one-site repulsion $U$.  We choose $t=1/4$ to establish a unit of energy
and choose $U=W=2$ which is sufficiently large that for $N_c\geq 4$, a Mott 
gap is present in the half-filled model. The phase diagram of the Hubbard 
model, calculated with DMFA, displays a range of behaviors including Mott 
insulating, antiferromagnetic and Fermi liquid regimes\cite{dmf_review}.  
The inclusion of non-local corrections yields significant changes to the 
phase diagram, including the enhancement of the Mott phase at half 
filling\cite{DCA_moukouri1}, the suppression of antiferromagnetism, the 
introduction of a d-wave superconducting phase and non-Fermi liquid 
behavior. These different regimes are delineated by calculating the 
antiferromagnetic (Fig.~\ref{Chi_AF}), bulk magnetic (Fig.~\ref{DOS_CHI}) 
and d-wave superconducting (Fig.~\ref{Pairing}) susceptibilities, and the 
single-particle self energy (Fig.~\ref{Sigmas_comp}).

\paragraph*{Antiferromagnetism}  

In Fig.~\ref{Chi_AF}, the inverse antiferromagnetic susceptibility is plotted 
versus temperature for dopings $\delta=0$ and $0.05$ and $N_c=4$ and $1$.  We 
plot the inverse susceptibility to show that in contrast to finite-sized system 
calculations, the susceptibility diverges for low doping at $T=T_N$. This 
indicates a transition to an antiferromagnetic phase.  In the DMFA for the 
2D model, or as we found previously for the infinite-dimensional 
model\cite{dmf_review}, the antiferromagnetism persists to relatively high 
temperatures and dopings.  The non-local dynamical fluctuations, included 
in the DCA for $N_c>1$, strongly suppress the antiferromagnetism.  Their 
effect becomes pronounced for low temperatures and dopings.  For example, 
when $\delta=0$, the $N_c=1$ and $N_c=4$ AF susceptibilities are identical 
at high temperatures due to the lack of non-local correlations, but separate 
as the temperature is lowered.  The $N_c=1$ susceptibility diverges with 
mean-field exponent of about one; whereas the $N_c=4$ result diverges at a 
much lower temperature with a larger exponent.  Consistent with the Mermin 
Wagner theorem, $T_N$ continues to fall for large values of $N_c$ (not shown). 

\paragraph*{The Pseudogap and Non-Fermi Liquid Behavior}

\begin{figure}[htb]
\epsfxsize=3.3in
\epsffile{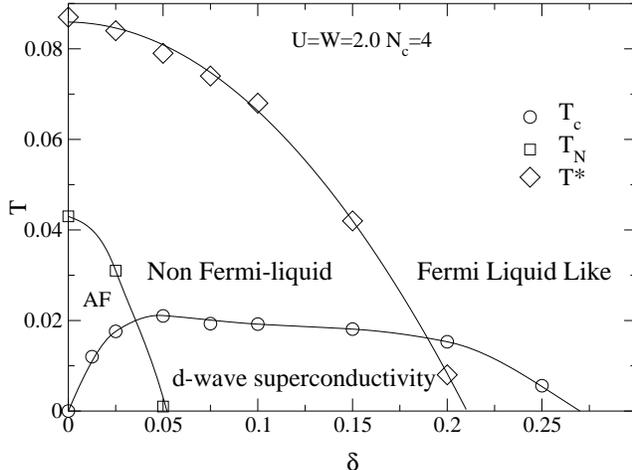}
\caption{The temperature-doping phase diagram of the 2D Hubbard model
calculated with QMC and DCA for $N_c=4$, $U=2$.  $T_N$ and $T_c$
were calculated from the divergences of the antiferromagnetic and
d-wave susceptibilities, respectively.  $T^*$ was calculated from
the peak of the bulk magnetic susceptibility.}
\label{temperatures}
\end{figure}

The bulk ($\k=0$) magnetic susceptibility and single-particle density of 
states (DOS) display evidence of a pseudogap for low doping $\delta < 0.2$. 
We show the bulk magnetic susceptibility in the inset to Fig.~\ref{DOS_CHI} 
for three different dopings.  For low to intermediate doping, it develops 
a peak at low temperatures, defining a temperature $T^*$.  
$T^* \alt T_N(N_c=1)$, the mean field transition temperature (see
Figs.~\ref{temperatures} and \ref{Chi_AF}). For $N_c>1$, it defines
the temperature where short-ranged spin correlations first emerge.
The underdoped bulk susceptibility data, $\delta\alt 0.075$, may be scaled onto 
one curve by plotting versus $T/T^*$ (not shown).   A similar peak or downturn 
and scaling is seen in the Knight-shift data of the cuprates\cite{htc_review}. 
The downturn of the susceptibility is accompanied by a loss 
of states near the Fermi energy. For temperatures $T < T^*$, a 
pseudogap begins to develop in the DOS, as 
shown in Fig.~\ref{DOS_CHI}.  The pseudogap is widest, measured from peak 
to peak, at low doping, and vanishes for $\delta \agt 0.2$.  The
depth of the pseudogap is greatest when $\delta\approx 0.05$, and it
vanishes as $\delta\to 0$, where it is replaced by a Mott gap of width 
$\approx U$.

The slow fall of $T^*$ with doping, indicates
that the short-ranged spin correlations diminish slowly upon doping. Thus, 
even in the region $T^* > T >T_N$ the antiferromagnetic correlations still 
have a significant effect.  This is supported by the behavior of the self 
energy for temperatures $T< T^*$. Once the DCA algorithm is converged, the 
lattice self energy is calculated by interpolating the cluster result on 
to the full lattice Brillouin zone.  Thus the lattice self energy at any 
$\k$ is dominated by the cluster self energy at the nearest cluster momentum. 
For a Fermi liquid, the self energy  
$\Sigma(\k,\omega) \sim (1-1/Z)\omega - i b \omega^2$ where $b>0$ and
$1/Z>1$.  Our results show that near half filling, the self energy displays 
non-Fermi liquid behavior.  This is illustrated in Fig~\ref{Sigmas_comp} 
where we plot the low-frequency self energy at the DCA cluster momenta for 
$\delta=0.05$ and $T=0.023$. For momentum points 
near $\k=(\pi,0)$, the imaginary part of the self energy crosses the Fermi 
energy almost linearly.  Concomitant with this behavior is a pseudogap 
of width $\approx |J| \approx 4t^2/U$ in the single-particle spectra 
$A(\k,\omega)$ for momenta near $\k=(\pi,0)$ (not shown).

The pseudogap and the anomalies in the self energy vanish when $T^*$ falls 
to zero. Here, as shown on the right of Fig.~\ref{Sigmas_comp}, the self 
energy becomes Fermi-liquid-like with quasiparticle weight $Z\approx 1/2$. 
A systematic study of the evolution of the single-particle spectra and the 
Fermi surface will be presented elsewhere.

It is important to stress that the pseudogap, the downturn of the bulk 
magnetic susceptibility and the non-Fermi liquid behavior in the self 
energy are {\it absent} when $N_c=1$ due to the lack of non-local 
fluctuations. 

\paragraph*{Superconductivity}  

We searched for many different types of superconductivity, including s, 
extended-s, p and d-wave, of both odd and even frequency and we looked 
for pairing at both the zone center and corner.  Of these, only the 
odd-frequency s-wave and even-frequency d-wave pair field susceptibilities 
at the zone center were strongly enhanced, and only the d-wave 
susceptibility diverged. This is illustrated in Fig.~\ref{Pairing} where 
the odd-frequency s-wave and even frequency d-wave $\q=0$ susceptibilities 
are plotted versus temperature for $U=2$ and $\delta=0.05$.  The s-wave 
and extended s-wave $\q=0$ even frequency susceptibilities are also 
plotted for comparison.

The behavior of the d-wave pair-field susceptibility as a function of
temperature for $N_c=1$ and $4$ and $\delta=0$ and $0.05$ is shown in 
the inset to Fig.~\ref{Pairing}. For $N_c=1$ there is no tendency 
towards pairing.  For the DMFA there is no pairing with symmetries 
lower than the lattice symmetry (i.e., p-, d-wave, etc.)\cite{DMF_jarrell}. 

d-wave pairing is strongly enhanced for $N_c=4$ over the corresponding DMFA 
results.  However, for $\delta=0$ the inverse susceptibility rises 
abruptly as the temperature is lowered and the Mott gap opens in the DOS.
The Mott gap becomes more pronounced as $N_c$ increases\cite{DCA_moukouri1}, 
so that for larger clusters the gap prevents superconductivity even for 
$U < W$.  If charge excitations are gapped, then pairing is suppressed.  
At half filling, for $U=2$ the gap is of order $U$, and is much larger 
than the magnetic exchange energy $|J|\sim 4t^2/U = 0.125$, so that the 
opening of the Mott gap will suppress any magnetically mediated pairing.  
Away from half filling the width of the pseudogap in the charge excitation 
spectrum is much smaller, on the order of $J$, so magnetically mediated 
pairing is possible.  For $N_c=4$ and $\delta=0.05$, the d-wave pair field 
susceptibility diverges at $T_c\approx 0.021$, with an exponent which is 
less than one, indicating that the fluctuations beyond DMFA which suppress 
the antiferromagnetism are also responsible for pairing.  

The phase diagram of the system is shown in Fig.~\ref{temperatures}.
In addition to the d-wave superconducting and antiferromagnetic
phase boundaries, we include $T^*$, the pseudogap temperature
fixed by the the peak bulk susceptibility.  At low temperatures, it 
serves as a boundary separating the observed Fermi liquid and 
non-Fermi liquid behavior.  For $T<T^*$ and $\delta<0.2$ the self energy 
shows non-Fermi liquid character for the parts of the Fermi surface 
closest to $\k=(\pi,0)$;  whereas, the low-temperature self energy 
is Fermi-liquid like for $\delta\agt 0.2$.  The d-wave transition 
temperature is maximum at $\delta\approx 0.05$.
The superconductivity persists to large doping, with $T_c$ dropping very 
slowly.  In contrast to experimental findings, the pairing instability 
(preceded by an AF instability) persists down to very low doping.  One 
possible reason for this is that the model remains very compressible down 
to very low doping $\delta\sim 0.025$. This could be due to the lack of 
long-ranged dynamical spin correlations or stripe formation which could 
become more relevant as $N_c$ increases or when multiple Hubbard planes 
are coupled together. 
The effect of such additional non-local corrections 
($N_c>4$) is presently unknown.  However, we believe that a finite 
mean-field coupling between Hubbard planes will stabilize the character 
of the phase diagram presented here as $N_c$ increases.

\paragraph*{Conclusion}
We have used QMC and DCA to study the effect of the initial non-local 
corrections on the phase diagram of the 2D Hubbard model at intermediate 
coupling $U=W$. The corrections make significant changes, including a 
strong suppression of the antiferromagnetism, the emergence of non-Fermi 
liquid (pseudogap) and d-wave superconducting regimes.  The critical 
exponent of the pair-field susceptibility is smaller than 
one, whereas the antiferromagnetic susceptibility diverges with a critical 
exponent larger than one. This indicates that the same fluctuations that 
suppress antiferromagnetism upon doping, mediate pairing. At half-filling 
the formation of the Mott gap of width $\gg J$ suppresses pairing.
%

\paragraph*{Acknowledgements} 
It is a pleasure to acknowledge useful discussions with
P.G.J.~van Dongen,
B.~Gyorffy,
D.~Hess,
C.~Huscroft,
J.~Keller,
H.R.~Krishnamurthy,
S.~Moukouri,
Th.~Pruschke,
J.~Zaanen
. This work was supported by NSF grants DMR-0073308 and PHY94-07194. 
Computer support was provided by the Ohio Supercomputer Center.

\end{document}